\documentclass[conference,compsoc,a4paper]{IEEEtran}
\usepackage[utf8]{inputenc}
\ifCLASSOPTIONcompsoc
  \usepackage[nocompress]{cite}
\else
  \usepackage{cite}
\fi
\usepackage{algorithmic}
\usepackage{algorithm}
\usepackage{tikz}
\usetikzlibrary{shapes.geometric, arrows.meta, positioning}
\usepackage{listings}
\usepackage{xcolor}
\usepackage{booktabs}
\usepackage{url}
\usepackage{microtype}

\begin{document}

\title{LLM2Alloy: Investigating LLM-Generated Formal Specifications\\
for Automated Test Derivation in Production Software}

\author{
\IEEEauthorblockN{Tasmim Rashid}
\IEEEauthorblockA{North Dakota State University\\
tasmim.rashid.dristy@gmail.com}
\and
\IEEEauthorblockN{Muhammad Zubair Malik}
\IEEEauthorblockA{North Dakota State University\\
zubair.malik@ndsu.edu}
}

\maketitle

\begin{abstract}
We present an exploratory study on using Large Language Models (LLMs)
to generate Alloy formal specifications from both requirements
documentation and production source code, and to derive executable
test cases from those specifications. We evaluate on two real
open-source Python libraries: Flipper, a feature flag management
system, and Cerberus, a data validation library.

In both cases, the LLM produced workable Alloy specifications and
executable tests without any manual correction. For Flipper, our
pipeline uncovered a genuine bug that the existing test suite had
missed: the library silently accepts duplicate flag names, directly
contradicting its  documented uniqueness requirement. A direct LLM
baseline---generating tests from the same README but skipping the
Alloy step---achieved 68\% branch coverage yet failed to catch
this bug across all three independent runs. This suggests that
introducing a formal intermediate representation can surface
constraint-level defects that coverage-oriented generation may miss.

For Cerberus, the code-derived specification captured an implicit
abstraction over sized types that the documentation-derived spec
omitted, producing two additional tests. Across both libraries,
code-based specifications showed lower variance in test generation
(mean SD~=~2.15) than documentation-based ones (mean SD~=~5.0),
though whether this generalises remains an open question.
\end{abstract}


\section{Introduction}

A persistent challenge in software testing is that tests are usually
written by the same developers who wrote the code. This means the
tests tend to reflect the developer's mental model of the
implementation, not necessarily what the requirements actually say.
Gaps between the two can quietly persist for a long time.

In this work, we explore whether Large Language Models (LLMs) can
help close that gap. Specifically, we ask whether an LLM can generate
formal specifications from both a system's requirements documentation
and its source code, and whether tests derived from those
specifications are meaningful enough to catch real bugs. We use Alloy
as our specification language because it is lightweight and precise,
and because its distinction between \texttt{pred} (a condition to
verify) and \texttt{fact} (an assumed invariant) turned out to be
diagnostically useful in our experiments.

The idea of using two independent sources---documentation and
code---is deliberate. If the two resulting specifications disagree
structurally, that disagreement is itself a signal: it may indicate
that what the system is documented to do and what it actually does
have drifted apart. We investigate this through two research questions:

\begin{itemize}
  \item \textbf{RQ1:} Can an LLM generate workable Alloy
  specifications from real production source code and requirements
  documentation?
  \item \textbf{RQ2:} Do specifications generated from documentation
  and from source code produce different or complementary results?
\end{itemize}

\section{Background and Related Work}

\subsection{Alloy and TestEra}
Alloy is a lightweight formal specification language based on
relational logic, introduced by Jackson~\cite{jackson2002}.
Marinov and Khurshid~\cite{marinov2001} showed that Alloy
specifications can automatically generate test inputs for Java
programs, but their TestEra framework required human experts to write
the specs. Our work removes that requirement by having an LLM generate
them automatically from existing documentation and source code.

\subsection{LLMs for Alloy Generation}
Hong et al.~\cite{hong2025} showed that LLMs can generate correct
Alloy formulas from natural language descriptions of abstract
properties such as DAGs and transitive relations. Their work validates
generated specs against reference formulas but stops short of
executing them against real software. We extend this by targeting
production libraries and running the resulting tests.

\subsection{LLM-Based Spec Repair}
Alhanahnah et al.~\cite{alhanahnah2024} investigated using LLMs to
repair broken Alloy specifications. Their work assumes a spec already
exists. We instead generate specs from scratch, starting only from
the two artefacts that are almost always present in any software
project: documentation and source code.

\subsection{Specification Generation in Other Formalisms}
Generating formal specifications with LLMs is not unique to Alloy.
SpecGen~\cite{specgen2025} demonstrates LLM-based generation of
method-level specifications for Java. Work on generating temporal
logic properties from code and natural language~\cite{wang2024}
further shows that LLMs can target a range of formal languages. Our
contribution is not the use of Alloy per se, but the \emph{dual-source}
strategy: generating independent specifications from both documentation
and code, then treating structural divergence between them as a signal
for requirement--implementation drift.

\subsection{Gap Addressed}
Prior work either stops at specification generation, uses toy
examples, or requires manual effort. We ask whether the full chain
from real production code and documentation, through LLM-generated
specs, to executable tests can work end to end, and whether the
tests are good enough to find bugs that other methods miss.

\section{Approach}

\subsection{Pipeline Overview}
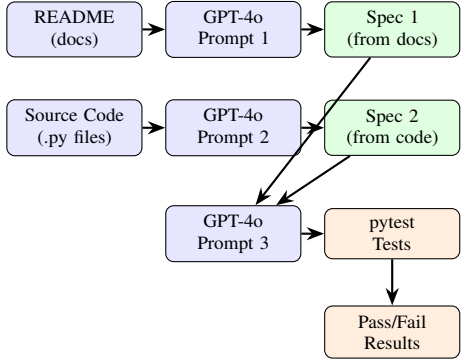
\begin{figure}[t]
\centering
\begin{tikzpicture}[
  ibox/.style={rectangle, rounded corners=3pt, draw, fill=blue!10,
               text width=1.55cm, align=center, minimum height=0.6cm,
               font=\scriptsize},
  sbox/.style={rectangle, rounded corners=3pt, draw, fill=green!12,
               text width=1.55cm, align=center, minimum height=0.6cm,
               font=\scriptsize},
  rbox/.style={rectangle, rounded corners=3pt, draw, fill=orange!13,
               text width=1.55cm, align=center, minimum height=0.6cm,
               font=\scriptsize},
  arr/.style={-Stealth, thick}
]
\node[ibox] (readme) at (0,    0)    {README\\(docs)};
\node[ibox] (src)    at (0,   -1.3)  {Source Code\\(.py files)};
\node[ibox] (llm1)   at (2.1,  0)    {GPT-4o\\Prompt 1};
\node[ibox] (llm2)   at (2.1, -1.3)  {GPT-4o\\Prompt 2};
\node[sbox] (spec1)  at (4.2,  0)    {Spec 1\\(from docs)};
\node[sbox] (spec2)  at (4.2, -1.3)  {Spec 2\\(from code)};
\node[ibox] (llm3)   at (2.1, -2.7)  {GPT-4o\\Prompt 3};
\node[rbox] (tests)  at (4.2, -2.7)  {pytest\\Tests};
\node[rbox] (res)    at (4.2, -4.0)  {Pass/Fail\\Results};
\draw[arr] (readme) -- (llm1);
\draw[arr] (src)    -- (llm2);
\draw[arr] (llm1)   -- (spec1);
\draw[arr] (llm2)   -- (spec2);
\draw[arr] (spec1)  -- (llm3);
\draw[arr] (spec2)  -- (llm3);
\draw[arr] (llm3)   -- (tests);
\draw[arr] (tests)  -- (res);
\end{tikzpicture}
\caption{LLM2Alloy pipeline. Documentation and source code are
independently translated into Alloy specs by GPT-4o; each spec
generates a pytest suite executed against the real library.}
\label{fig:pipeline}
\end{figure}

Our pipeline has three LLM-driven stages, shown in
Fig.~\ref{fig:pipeline}. We extracted content from each repository
manually and submitted it to GPT-4o via the web interface in fresh
chat sessions, with no API calls, to avoid any cross-contamination
between runs.

\begin{enumerate}
  \item Extract README documentation $\rightarrow$ submit to GPT-4o
        with Prompt~1 $\rightarrow$ obtain Alloy Spec~1
  \item Extract source code $\rightarrow$ submit to GPT-4o
        with Prompt~2 $\rightarrow$ obtain Alloy Spec~2
  \item Submit each Alloy spec with Prompt~3 $\rightarrow$ obtain
        a pytest test suite
  \item Run the tests with pytest $\rightarrow$ record pass/fail
\end{enumerate}

\subsection{Why Alloy as an Intermediate Step}
\label{sec:whyalloy}
One could imagine asking the LLM to generate tests directly from the
README or source code, skipping the Alloy step. We evaluate that
approach as a baseline in Section~\ref{sec:baseline}. But there are
reasons to think an intermediate formal representation helps.

First, Alloy's relational logic is well suited to the kinds of
set-membership and uniqueness constraints that matter for
specification-level bugs. Second, Alloy's \texttt{pred}/\texttt{fact}
distinction carries semantic weight: a \texttt{pred} is something to
be checked, while a \texttt{fact} is assumed unconditionally true.
When the documentation-derived and code-derived specs disagree on
which one to use for the same property, that disagreement is
informative. Third, prior work~\cite{hong2025} has shown LLMs can
generate valid Alloy from natural language, so the generation step
itself is on solid ground.

\subsection{Automation Level}
The pipeline is fully LLM-driven. We made no manual corrections to
any generated specification or test. The only human judgement involved
was deciding which README sections and source files to include in
each prompt.

\subsection{Prompt Design}
All three prompts were minimal: they named the target language or
format and asked for output only, with no tutorials or examples. GPT-4o
produced syntactically valid Alloy in every run without correction,
which suggests the model has enough training-time exposure to Alloy to
work from a brief role-description alone.

\textbf{Prompt 1 --- README to Alloy spec:}
\begin{lstlisting}
You are an expert in Alloy formal specification language.
Generate an Alloy specification for [system name] based
on these requirements: [README content]
Output ONLY valid Alloy code, no explanations.
\end{lstlisting}

\textbf{Prompt 2 --- Source code to Alloy spec:}
\begin{lstlisting}
You are an expert in Alloy formal specification language.
Reverse-engineer an Alloy specification from this code:
[source code]
Output ONLY valid Alloy code, no explanations.
\end{lstlisting}

\textbf{Prompt 3 --- Alloy spec to tests:}
\begin{lstlisting}
Here is an Alloy specification. Generate Python pytest
tests that validate ONLY the constraints defined in this
spec against the real library.
Output ONLY pytest code, no explanations. [Alloy spec]
\end{lstlisting}

\subsection{Replication Protocol}
Each prompt was submitted three times in separate sessions to assess
LLM non-determinism. We did not constrain the number of generated
tests; we wanted to observe what the model produces naturally from
each spec. Constraining the count is a simple extension left for
future work.

\section{Case Study}

\subsection{Subject Applications}
We chose two mature open-source Python libraries with real production
users, clear and localised validation logic, and constraints that
translate naturally into Alloy. This is a proof-of-concept study, not
a statistical one.

\begin{itemize}
  \item \textbf{Flipper}~\cite{flipper}
  \item \textbf{Cerberus}~\cite{cerberus}
\end{itemize}

\subsection{Flipper: Feature Flag Management}

\textbf{Context.} Flipper is a Python client for feature flag
management, used in production by over 30,000 companies.

\textbf{Source code extracted.} The \texttt{create()} method in
\texttt{flipper/client.py}. It never calls \texttt{exists()} before
writing to the store---which is where the bug lives.

\begin{lstlisting}[language=Python]
def create(self, feature_name, is_enabled=False,
           client_data=None):
    meta = Meta(created_date=unix_time())
    feature = Feature(name=feature_name, meta=meta)
    self._store.create(feature_name, feature.to_dict())
    if is_enabled:
        self.enable(feature_name)
    return FeatureFlagClient.wrap(feature, self)
\end{lstlisting}

\textbf{Documentation.} The README states:
\textit{``Each feature flag is identified by a unique name.''}

\textbf{Generated Alloy spec (Spec~1 --- from README):}
\begin{lstlisting}
sig FeatureFlag { name: one String, enabled: one Bool }
sig State { flags: set FeatureFlag }
pred UniqueNames[s: State] {
  all disj f1, f2: s.flags | f1.name != f2.name
}
pred Create[s, s': State, n: String] {
  not Exists[s, n]
  some f: FeatureFlag |
    f.name = n and s'.flags = s.flags + f
}
pred ValidState[s: State] { UniqueNames[s] }
\end{lstlisting}

\textbf{Spec~1 interpretation.} \texttt{UniqueNames} is a
\texttt{pred}---something that may or may not hold, and should be
checked. \texttt{Create} explicitly guards against inserting a name
that already exists.

\textbf{Generated Alloy spec (Spec~2 --- from source code):}
\begin{lstlisting}
fact UniqueNames {
  all s: State |
    all disj f1, f2: s.flags | f1.name != f2.name
}
\end{lstlisting}

\textbf{Spec~2 interpretation.} Here uniqueness is a \texttt{fact}
assumed unconditionally true and therefore never tested. The LLM
reverse-engineered from code that never checks for duplicates, so
it encoded uniqueness as a background assumption rather than a
verifiable condition. The \texttt{pred}/\texttt{fact} divergence
between the two specs was the first sign something was wrong.

\subsection{Micro Ablation: Comparing Both Approaches on Flipper}

\begin{table}[t]
\caption{Flipper micro ablation. Mean and SD are over three
independent runs. Failures counts tests that exposed a constraint
violation. Bug Found indicates whether the duplicate-name bug
was detected.}
\begin{center}
\begin{tabular}{|c|c|c|c|c|}
\hline
\textbf{Approach} & \textbf{Mean} & \textbf{SD} &
\textbf{Failures} & \textbf{Bug Found} \\
\hline
Requirements (Spec 1) & 8.0 & 2.6 & 1 & Yes \\
\hline
Code (Spec 2) & 6.7 & 0.5 & 0 & No \\
\hline
\end{tabular}
\label{tab_ablation}
\end{center}
\end{table}

The documentation-based spec caught the bug because it treated
uniqueness as something to verify. The code-based spec assumed it was
always true and never tested it---a good illustration of how specs
derived from code can inadvertently inherit the same blind spots as
the code itself.

\subsection{Baseline Comparison: Does the Alloy Step Help?}
\label{sec:baseline}

To check whether the Alloy intermediate step is actually doing useful
work, we ran a direct baseline: generate pytest tests from the README
using a prompt comparable in specificity to Prompts~1 and~3 combined,
but with no Alloy step in between.

Across three runs, the baseline produced 37 tests consistently, all
of which passed, and achieved 68\% branch coverage (measured with
\texttt{pytest-cov}). Not a single test touched the uniqueness
constraint, and the duplicate-name bug went undetected every time.

The Alloy pipeline, by contrast, produced this test in all three runs:

\begin{lstlisting}[language=Python]
def test_create_duplicate_name_fails(self):
    client = FeatureFlagClient(MemoryFeatureFlagStore())
    client.create("duplicate_feature")
    with pytest.raises(Exception):
        client.create("duplicate_feature")
# FAILED: DID NOT RAISE <class 'Exception'>
\end{lstlisting}

The difference is that Alloy forced the uniqueness requirement into an
explicit, testable predicate. Direct generation, even at 68\% coverage,
never surfaced it. This points to a broader issue: coverage measures
how much code gets executed, but says nothing about whether the right
semantic constraints are being checked.

\textbf{Proposed fix:}
\begin{lstlisting}[language=Python]
def create(self, feature_name, is_enabled=False,
           client_data=None):
    if self.exists(feature_name):
        raise ValueError(
            f"Feature '{feature_name}' already exists.")
    # rest of method unchanged
\end{lstlisting}

\subsection{Cerberus: Data Validation}

\textbf{Context.} Cerberus validates Python dictionaries against
user-defined schemas. It has $\sim$3,100 GitHub stars and $\sim$2
million PyPI downloads per month.

\textbf{Key spec difference.} Spec~1 modelled strings and lists as
separate types. Spec~2 unified them under a \texttt{SizedVal}
abstraction, reflecting Python's \texttt{Iterable} interface:

\begin{lstlisting}
pred ValidateMinLength[f: Field] {
  (some f.r.minlength and f.value in SizedVal) implies
    (f.value.(SizedVal.size) >= f.r.minlength)
}
\end{lstlisting}

Because Spec~2 modelled the abstraction more faithfully, it generated
two additional tests covering list-length constraints that Spec~1
missed. All 34 Cerberus tests passed. This is a clean answer to RQ2:
the two source types produce genuinely complementary specifications.

\subsection{Prompt Stability Results}

\begin{table}[t]
\caption{Test generation stability across three independent runs.
R1--R3 are test counts per run; Mean and SD capture variability
from LLM non-determinism.}
\begin{center}
\begin{tabular}{|c|c|c|c|c|c|}
\hline
\textbf{Prompt} & \textbf{R1} & \textbf{R2} &
\textbf{R3} & \textbf{Mean} & \textbf{SD} \\
\hline
Flipper Req.  & 11 & 6  & 7  & 8.0  & 2.6 \\
\hline
Flipper Code  & 7  & 6  & 7  & 6.7  & 0.5 \\
\hline
Cerberus Req. & 16 & 18 & 4  & 12.7 & 7.4 \\
\hline
Cerberus Code & 18 & 22 & 13 & 17.7 & 3.8 \\
\hline
\end{tabular}
\label{tab_stability}
\end{center}
\end{table}

Code-driven prompts were noticeably more stable (mean SD~=~2.15) than
documentation-driven ones (mean SD~=~5.0). This is probably because
source code expresses constraints unambiguously, while natural language
documentation leaves more room for the model to interpret things
differently across runs. Importantly, the Flipper bug was detected in
all three README-based runs, so the finding is not a fluke of
non-determinism.

\section{Discussion}

\subsection{What Makes the Dual-Source Strategy Different}
The long-term goal of this work is to close the semantic gap between
what a system is supposed to do and what it actually does. Generating
specifications independently from documentation and code is the
mechanism we use to surface that gap. The key insight from the Flipper
case is that the \texttt{pred}/\texttt{fact} divergence between the
two specs was a direct structural fingerprint of the bug: one spec
said ``check this,'' while the other said ``assume this.'' That signal
does not emerge from generating a single specification, and it did not
emerge from direct test generation either.

This strategy is not tied to Alloy specifically. Other formal
languages that distinguish verifiable conditions from assumed
invariants could serve the same purpose. We chose Alloy partly because
prior work~\cite{hong2025} establishes that LLMs can generate it
correctly, and partly because its semantics made the divergence easy
to interpret.

\subsection{On Coverage as a Metric}
We recognise that test count is a weak metric---multiple tests can
cover the same lines. We include branch coverage for the baseline
(68\%) as a more informative comparison point. But even coverage has
limits here. The Flipper bug manifests as a \emph{missing guard}:
the code path that should reject a duplicate never exists, so no
test can cover it by accident. The only way to catch it is to
explicitly test the constraint. That is what the Alloy intermediate
step pushed the model to do. Future work should adopt mutation score
as a more sensitive metric for this class of defect.

\subsection{External Validation}
The duplicate-name bug has not yet been submitted as a pull request
to the Flipper repository. We plan to do so and report on the
maintainers' response in follow-up work.

\subsection{Threats to Validity}

\textbf{Internal validity.} Each prompt ran in a fresh session with
no shared context. Nothing was manually corrected. The main threat is
LLM non-determinism, which we address by replicating each prompt
three times.

\textbf{External validity.} Two libraries are enough for a
proof-of-concept, but we cannot claim that the results generalise. Other
libraries, languages, and LLMs may behave very differently.

\textbf{Construct validity.} Test count is an imperfect proxy for
quality. We use pass/fail outcomes and coverage as complementary
signals, and recommend mutation score for future work.

\section{Conclusion}

We set out to answer two questions: can an LLM generate workable Alloy
specifications from real production software, and do the two
sources---documentation and code---produce different, useful results?

Both answers are yes. For RQ1, GPT-4o produced syntactically valid
Alloy and executable tests in every run, with no manual corrections.
For RQ2, the two sources were genuinely complementary. The
documentation-derived spec caught a real bug in Flipper that the
existing test suite and a direct LLM baseline at 68\% branch coverage
both missed. The code-derived spec captured an abstraction in Cerberus
that the documentation omitted, adding test coverage the other
approach lacked.

More broadly, these results suggest that introducing a formal
intermediate representation can shift LLM-based testing from
surface-level pattern generation toward constraint-driven validation.
Whether that shift holds across a wider range of systems is the
natural next question.

Future work includes testing more repositories, comparing LLMs,
running more replication trials, integrating the Alloy Analyzer for
automated counterexample generation, and adopting coverage and
mutation score as standard evaluation metrics.

\ifCLASSOPTIONcompsoc
  \section*{Acknowledgments}
\else
  \section*{Acknowledgment}
\fi
The authors would like to thank the Overleaf built-in AI assistant
and Gemini for linguistic and \LaTeX-related suggestions.

\end{document}